\documentclass[aps,pra,twocolumn,superscriptaddress,longbibliography,10pt]{revtex4-1}

\usepackage{graphicx}
\usepackage{dcolumn}
\usepackage{bm}
\usepackage{times}
\usepackage{amsmath}
\usepackage{bbold}
\usepackage{epstopdf}
\usepackage{upgreek}
\usepackage[english]{babel}
\usepackage{color}
\usepackage[normalem]{ulem}
\usepackage[colorlinks]{hyperref}
\usepackage[colorinlistoftodos,prependcaption,textsize=tiny]{todonotes}
\hypersetup{
colorlinks=true,
linkcolor=blue,          
citecolor=blue,          
filecolor=magenta,       
urlcolor=cyan
}
\usepackage{xfrac}

\DeclareFontFamily{U}{mathx}{\hyphenchar\font45}
\DeclareFontShape{U}{mathx}{m}{n}{<-> mathx10}{}
\DeclareSymbolFont{mathx}{U}{mathx}{m}{n}
\DeclareMathAccent{\widebar}{0}{mathx}{"73}

\begin{document}


\title{Selection rules for cavity-enhanced Brillouin light scattering from magnetostatic modes}

\author{J. A. Haigh}
\email{jh877@cam.ac.uk}
\affiliation{Hitachi Cambridge Laboratory, Cambridge, CB3 0HE, UK}
\author{N. J. Lambert}
\affiliation{Cavendish Laboratory, University of Cambridge, Cambridge, CB3 0HE, UK}
\author{S. Sharma}
\affiliation{Kavli Institute of NanoScience, Delft University of Technology, 2628 CJ Delft, The Netherlands}  
\author{Y. M. Blanter}
\affiliation{Kavli Institute of NanoScience, Delft University of Technology, 2628 CJ Delft, The Netherlands}  
\author{G. E. W. Bauer}
\affiliation{Institute for Materials Research \& WPI-AIMR \& CSRN, Tohoku University, Sendai 980-8577, Japan} 
\affiliation{Kavli Institute of NanoScience, Delft University of Technology, 2628 CJ Delft, The Netherlands}  
\author{A. J. Ramsay}
\affiliation{Hitachi Cambridge Laboratory, Cambridge, CB3 0HE, UK}

\date{\today}

\begin{abstract}
We experimentally identify the magnetostatic modes active for Brillouin light scattering in the optical whispering gallery modes of a yttrium iron garnet sphere. Each mode is identified by magnetic field dispersion of ferromagnetic-resonance spectroscopy and coupling strength to the known field distribution of the microwave drive antenna. Our optical measurements confirm recent predictions that higher-order magnetostatic modes can also generate optical scattering, according to the selection rules derived from the axial symmetry. From this we summarize the selection rules for Brillouin light scattering. We give experimental evidence that the optomagnonic coupling to non-uniform magnons can be higher than that of the uniform Kittel mode.
\end{abstract}

\maketitle

\section{Introduction}

Brillouin light scattering (BLS) is an important technique for the study of magnons \cite{chumak_magnon_2015}, the elementary excitations of the magnetic order in ferromagnets \cite{demokritov_brillouin_2001} and antiferromagnets \cite{fleury_two-magnon_1967}. The energy and wavevector sensitivity of the technique allows, for example, mapping of dispersion relations \cite{di_direct_2015}, including with spatial resolution \cite{demokritov_micro-brillouin_2008}. As an experimental tool, BLS is typically used as a probe of magnetization dynamics, which are often excited by some other external stimulus. 

In contrast, there is recent interest in taking magnon BLS to a new regime, in which the optical fields and magnetization dynamics are sufficiently strongly coupled that they cannot be treated independently. This is akin to the strong parametric coupling limit in optomechanics \cite{aspelmeyer_cavity_2014}, but with the mechanical harmonic oscillator replaced by a magnetic one. While strong coupling of magnons to GHz microwave-cavity photons is readily achieved \cite{tabuchi_hybridizing_2014,zhang_strongly_2014,roberts_magnetodynamic_1962}, similar levels of coupling to optical photons is more difficult to achieve. Efforts have focused on enhancing BLS in magneto-optical resonators \cite{osada_cavity_2016,zhang_optomagnonic_2016,haigh_triple-resonant_2016}, exploiting the highly confined optical whispering gallery mode (WGM) resonances of polished ferrimagnetic yttrium iron garnet spheres \cite{haigh_magneto-optical_2015}. In these experiments, a BLS enhancement is indeed observed \cite{haigh_triple-resonant_2016}, and theoretical calculations show that, in a different geometry, strong coupling can in principle be reached \cite{viola_kusminskiy_coupled_2016}.

\begin{figure}%
\includegraphics[width=\columnwidth]{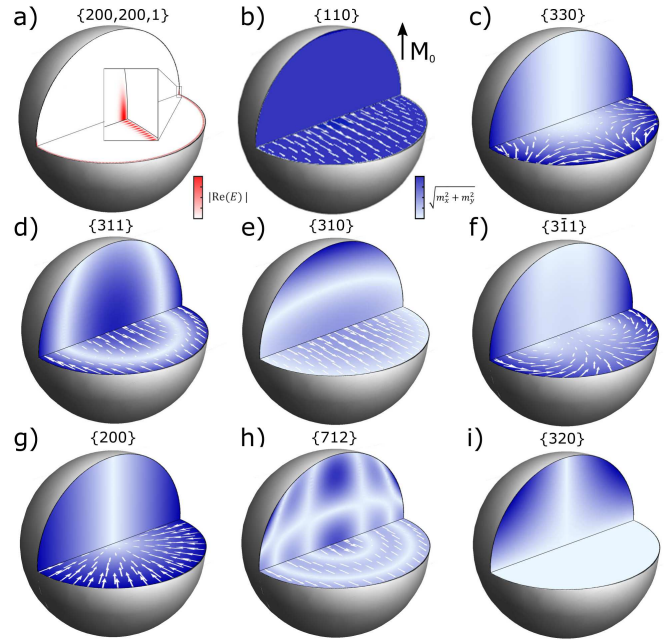}%
\caption{Spatial variation of (a) an optical WGM and (b-i) dynamic magnetization modes of a yig sphere. The labels above the spheres are the angular momentum and radial mode numbers $\{l,m,q\}$ (with subscript $m$ for magnetic modes, see text). (a) The in-phase intensity of the electric field for an optical WGM with $l=m=200$ (in the actual experiment, $l,m\sim1000$). (b-i) The color indicates the intensity of the dynamic magnetization, while arrows indicate the in-phase direction. Negative mode indices are indicated by an over-bar. The static magnetization $\bf{M}_0$ is indicated in (b).}%
\label{modes}%
\end{figure}

Experiments so far have focused on the Kittel mode, in which the magnetization precesses uniformly across the entire sphere. However, a recent theoretical analysis \cite{Sharma_light_2017} showed that the enhanced BLS should be large for some spatially varying magnetic modes as well. Due to better spatial overlap with WGMs, these could be expected to have stronger optomagnonic coupling.

Here, we report experiments to detect and identify the BLS active magnetostatic modes in YIG spheres. These modes can be specified by selection rules given by the axial symmetry of the optical and magnetic modes involved in the scattering.

The modes of interest in this paper, both optical and magnetic, are defined by the symmetry of the yttrium iron garnet sphere. The optical WGMs are specified by a set of indices $\{l,m,q\}$ and $\sigma$. These indices give the number of radial $(q-1)$, azimuthal $m$, and polar $(l-m)$ nodes in the electric field \cite{schunk_identifying_2014}. The linear polarization $\sigma$ of the WGM is either horizontal $h$ or vertical $v$ w.r.t. the WGM orbit plane, also known as transverse magnetic (TM) and transverse electric (TE), respectively. There is a frequency splitting between $h$ and $v$ polarized modes due to the symmetry breaking associated with the interface at the surface of the sphere. Due to angular momentum conservation, BLS from magnons is forbidden between modes with equal polarization \cite{le_gall_theory_1971}. Therefore, as the WGMs modes are linearly polarized, scattering is always between orthogonal polarizations $h\leftrightarrow v$. The optical modes with large $l,|m|\gg 1$ are localized in the $x,y$ plane at the equator and form the WGMs, as shown in Fig.\,\ref{modes}(a).

The magnetostatic modes of the YIG sphere correspond to normal modes of the small dynamic component of the magnetization, transverse to the large static magnetization component along the static magnetic field direction. The mode forms are governed by dipolar interaction, and can be analytically calculated in the magnetostatic limit \cite{walker_resonant_1958,fletcher_ferrimagnetic_1959}, where in addition the exchange energy is neglected due to the long magnon wavelengths relative the exchange length. They can be labeled by three indices $\{l_{m},m_{m},q_{m}\}$ where $m_{m}$ is the number of azimuthal nodes in the tangential component of the magnetization \footnote{$\{n,m,r\}$ \ in Ref.\thinspace \cite{fletcher_ferrimagnetic_1959}.}.

Fig.\,\ref{modes}(b) shows the node-less Kittel mode. The other modes have additional nodal planes in the form of ellipsoids whose number and ellipticity is governed by $l_{m}$ and $q_{m}$, respectively and are shown in Fig.\,\ref{modes}(c-i). In microwave experiments, strong coupling of several higher order modes to microwave resonators has been achieved \cite{goryachev_high-cooperativity_2014,lambert_identification_2015,karenowska_strong_2017}, despite the fact that only the Kittel mode has any net dynamic magnetization. This is possible due to inhomogeneity in either the microwave or the applied static magnetic field. In our experiments, we exploit both to allow us to drive various non-uniform magneto-static modes.

As the wavelength $\lambda\approx1310$\,nm of the light is much smaller than the radius of the sphere,  $a = 0.5$\,mm, the WGMs occupy a very narrow band around the equator (Fig.\,\ref{modes}(a)). The scattering intensity therefore depends on (1) the intensity of the magnetostatic modes at the equator, and (2) the texture of the dynamic magnetization along the WGM path. The small proportion of the sphere occupied by the optical WGM is the primary reason for the low optomagnetic coupling strength of the Kittel mode. 

\section{Experimental setup}

\begin{figure}%
\includegraphics[width=\columnwidth]{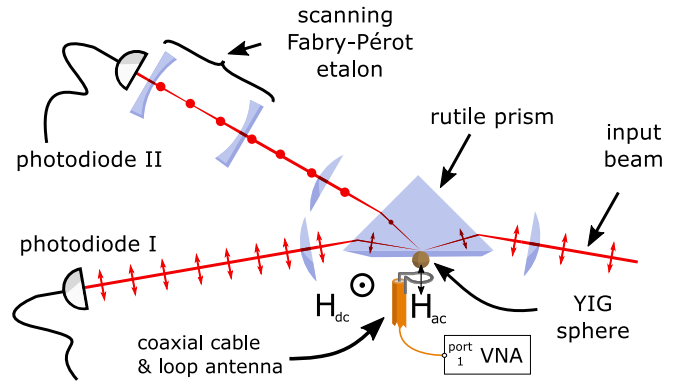}%
\caption{Experimental setup. Linearly polarized input light is evanescently coupled into the YIG sphere via a rutile prism. Photodiode I measures the transmitted input beam, in order to identify the WGM resonances. Photodiode II measures the polarization rotated output from the YIG sphere, with a scanning Fabry-P\'erot etalon in the beam path providing spectral resolution. A permanent NdFeB magnet saturates the magnetization perpendicular to the WGM orbit. The microwave measurements are made with the loop antenna using a vector network analyzer (VNA).}%
\label{setup}%
\end{figure}

A schematic of the experimental setup is shown in Fig.\,\ref{setup}. A rutile prism coupler is used to optically excite the whispering gallery modes, while the magnetostatic modes are driven by a small loop antenna. The 1\,mm diameter YIG sphere is mounted on a ceramic rod.

The WGMs are probed with a tunable external-cavity diode laser with linewidth $\approx$1\,MHz. Due to the birefringence of the coupling prism, the reflected linearly polarized input beam and the polarization-rotated scattered beam are spatially separated and can be measured independently. The reflected beam is measured on a photodiode and is used to identify the WGMs. The polarization scattered light is passed through a scanning Fabry-P\'{e}rot etalon to spectrally resolve the BLS. Whilst in previous experiments \cite{haigh_magneto-optical_2015,haigh_triple-resonant_2016} we have studied both input polarizations, here we focus solely on measurements for $h$-input (TM) polarization, where better out-coupling of the BLS light from the birefringent coupling prism is achieved.

\begin{figure*}%
\includegraphics[width=\textwidth]{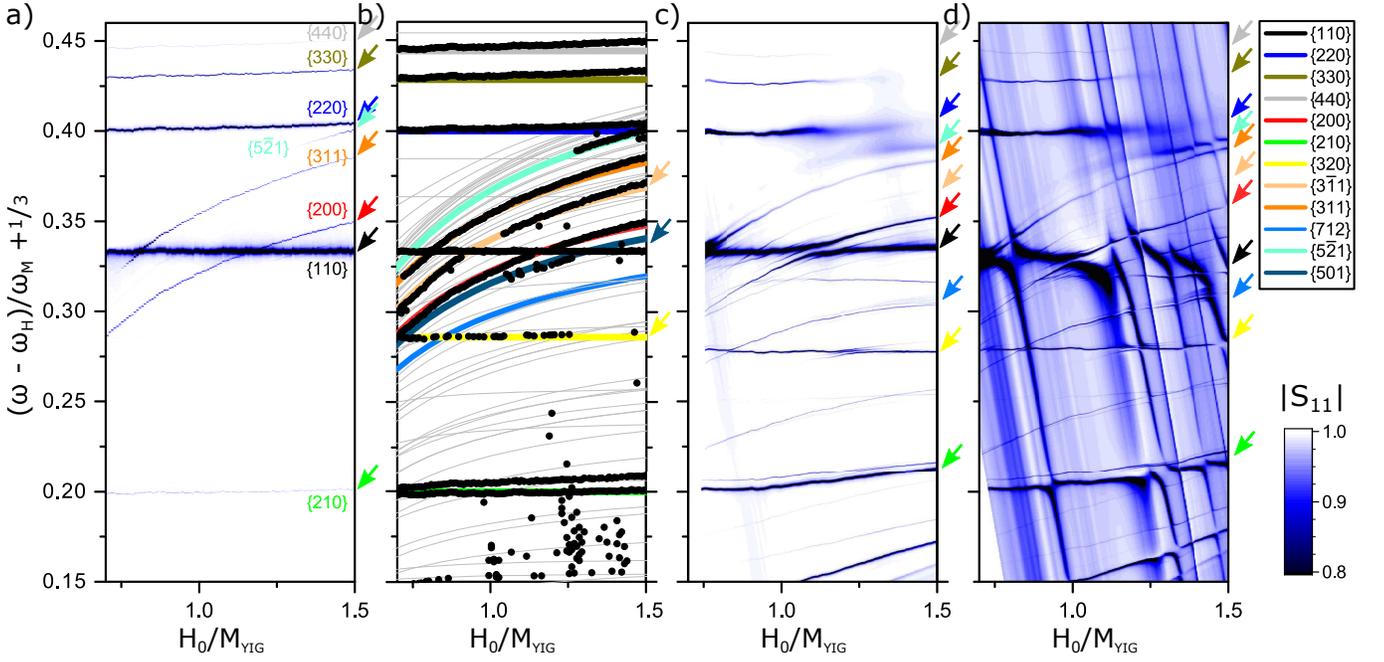}%
\caption{Identification of magnetostatic modes from magnetic field dispersion. (a) Microwave reflection coefficient $|S_{11}|$ as a function of magnetic field and reduced frequency $(\omega-\omega_{H})/\omega_{M}+\sfrac{1}{3}$ measured in uniform applied magnetic field generated by an electromagnet. (b) Calculated eigenfrequencies of magnetostatic modes with indices $\{l_{m},m_{m},q_{m}\}$, overlaid with dips from (a). This is used to identify the magnons excited by the microwave antenna. (c) Same as (a) but in non-uniform magnetic field of a permanent magnet showing normal mode splittings and additional magnons. (d) Same as (c), but with rutile coupling prism in place indicating the microwave modes of the prism alongside that of the YIG sphere.}%
\label{msm}%
\end{figure*}

For microwave characterization of the magnetostatic modes, we measure absorption dips in the reflection coefficient $|S_{11}|$ of the loop antenna with a vector network analyzer. The static magnetic field is applied using a permanent NdFeB magnet. The magnitude of the magnetic field can be controlled by shifting the position of the magnet relative to the YIG sphere.

\section{Expected modes active for BLS} \label{Sec:Theory}

We briefly review the modes expected to be active for BLS in our experimental setup using recent theoretical calculations of the optomagnonic coupling constants \cite{Sharma_light_2017}.

A photon in a $\sigma _{i}=h$ polarized input WGM $\{l_{i},m_{i},q_{i}\}$ can undergo anti-Stokes scattering by a magnon $\{l_{m},m_{m},q_{m}\}$ into a $\sigma_{o}=v$ polarized output WGM $\{l_{o},m_{o},q_{o}\}$, while Stokes scattering is strongly suppressed \cite{haigh_triple-resonant_2016,Sharma_light_2017}. The constraints on the coupling constant $G$ can be summarized as \cite{Sharma_light_2017}, 
\begin{equation}
G \propto \delta_{q_{i},q_{o}}\left\langle l_{i},m_{i};l_{m},m_{m}\middle|l_{o},m_{o}\right\rangle.  \label{SelectionRules}
\end{equation}
This expression effectively captures the mode matching between the three modes. The first factor gives a radial selection rule, $q_{i} = q_{o}$. The second factor is the Clebsch-Gordan coefficient governing the angular momentum conservation.

For WGMs, $m_{i} \approx l_{i}$ and $m_{o} \approx l_{o}$, while for the magnons excited by microwaves, $l_{m},m_{m} \sim 1 \ll l_{i},l_{o} \sim 10^3$. Under such conditions, the optical interaction with the magnon occurs only in the thin band occupied by the WGMs near the equator. The long-wavelength nature of magnons therefore preserves the transverse field distribution of WGMs. This gives the radial selection rule above, and considering polar direction, that $l_{o} - m_{o} = l_{i} - m_{i}$. The wave-matching conditions in the azimuthal direction dictates $m_{o} = m_{m} + m_{i}$. This implies that $G$ is approximately zero unless $l_{o} - l_{i} = m_{m}$, which is confirmed by explicit calculation of the Clebsch-Gordan coefficient. BLS scatters photons into the mode given by $\{l_{i} + m_{m}, m_{i} + m_{m},q_{i}\}$ which is fixed by the incident WGM and the magnon.

For significant coupling we require a non-zero magnon density at the equator, where WGMs reside. From explicit solutions \cite{fletcher_ferrimagnetic_1959}, the magnetostatic mode amplitudes vanish at equator for odd $l_{m} - m_{m}$ (see Fig.\,\ref{modes}(b-i)).

Finally, we consider the energy conservation. The $l_{o}=l_{i}+1$, $l_{o}=l_{i}$ and $l_{o}=l_{i}-1$ transitions have frequencies of $7$\,GHz, $40$\,GHz, and $50$\,GHz, respectively,  fixed by the optical cavity free spectral range and geometrical birefringence. The linewidth of the WGM of $\approx1$\,GHz is much smaller than the frequency spacing between these transitions, ensuring the selectivity of the resonance condition. In our setup the maximum field is $\approx300$\,mT, corresponding to a ferromagnetic resonance frequency $\approx8.5$\,GHz. Hence, only the  $l_{o}=l_{i}+1$ transitions are observed \cite{haigh_triple-resonant_2016}. Comparing this resonance condition to $l_{o} - l_{i} = m_{m}$ derived previously, we therefore have $m_{m} = 1$. The fact that $l_{m} - m_{m}$ must be even then restricts $l_{m}$ to be an odd integer. While it is more difficult to couple microwaves to high $l_{m}$ modes, increasing $l_{m}$ typically increases the equatorial magnon density and hence, is likely to have higher optomagnonic coupling. We note that the equatorial magnon density also depends on $q_{m}$, but its discussion is beyond the scope of this work.

In summary, the magnons expected to be active for BLS should have $l_{m} = 1,3,5,\dots$ and $m_{m} = 1$. Note that the sign of $m_{m}$ in the allowed transitions is for the magnetic field direction shown in Fig.\,\ref{setup}, such that the angular momentum of the WGMs is parallel to the static magnetization. Similar arguments show that for the opposite magnetic field (or WGM circulation direction), the expected magnons should have $l_{m} = 1,3,5,\dots$ and $m_{m} = -1$. For opposite input polarization, energy conservation leads to preferential Stokes scattering, but the same selection rules apply.

\section{Identification of magnetostatic modes} \label{Sec:MW_Exp}

The microwave reflection coefficient $|S_{11}|$ of the loop antenna \cite{klingler_gilbert_2017} is measured in the experimental setup shown in Fig.\thinspace \ref{setup}. Two complications hinder the labeling of the magnetostatic mode spectra: (1) the inhomegeniety of the static magnetic field of the permanent magnet. (2) The rutile coupling prism is a good microwave dielectric resonator that interferes with the magnetic resonance. Therefore, we first carry out a simpler experiment by transferring the loop antenna with YIG sphere (without the prism) into a separate electromagnet with uniform static magnetic field. The results are summarized in Fig.\thinspace \ref{msm}(a). We follow Ref.\,\onlinecite{fletcher_ferrimagnetic_1959} and plot the reduced frequency  $\left( \omega -\omega _{H}\right) /\omega _{M}+\sfrac{1}{3}$, where the Larmor frequency $\omega_H=\gamma \mu_0 H_0$ is subtracted so that the dispersion can be seen more clearly. Here, the gyromagnetic ratio $\gamma=28$\,GHz/T and $\omega_{M} = \gamma \mu_0 M_\textsc{yig}$, with $\mu_0 M_\textsc{yig} = 180$\,mT. We use the $\{110\}$ (Kittel) and $\{220\}$ modes as magnetic field sensors, aligning them to their expected position in reduced frequency. These can be identified by their frequency separation $\omega_{M}/15$, which is independent of magnetic field and depends only on the saturation magnetization. The rescaled map of the observed microwave reflection coefficient $|S_{11}|$ is shown in Fig.\thinspace \ref{msm}(a).

The positions of the resonances in Fig.\,\ref{msm}(a) are plotted in Fig.\,\ref{msm}(b), along with expected mode frequencies \cite{fletcher_ferrimagnetic_1959}. There is clear agreement with several sets of points indicating that several non-Kittel modes are driven by the loop antenna (highlighted by colored lines). If the drive field distribution of this antenna were uniform, only the Kittel mode would couple to the microwave line. However, non-uniformity in the drive field allows other magnetostatic modes to be driven as well. 

To help identify the observed magnetostatic modes, we numerically calculated the magnon mode overlap with the drive field distribution of the loop antenna treated as a current loop. All the modes labeled in Fig.\,\ref{msm}(a) have microwave coupling strength greater than 0.1\% of the Kittel mode, apart from the $\{521\}$ and $\{210\}$ modes which are much weaker in the model. For example, the relative microwave coupling strength for the $\{200\}$ mode is estimated to be $\approx$4\%. 

Next, we transfer the YIG sphere and microwave antenna to the optical setup (with rutile prism removed) in which the static magnetic field is generated by a small permanent magnet since there is no room for an electromagnet. The differences between the measured $|S_{11}|$ in Fig.\thinspace \ref{msm}(c) and (a) are caused by the inhomogeneous dc magnetic field. We again use the $\{110\}$ and $\{220\}$ modes as sensors for the magnetic field distribution, which can be estimated by analytical expressions for a cuboid magnet \cite{engel-herbert_calculation_2005}.

\begin{figure}%
\includegraphics[width=\columnwidth]{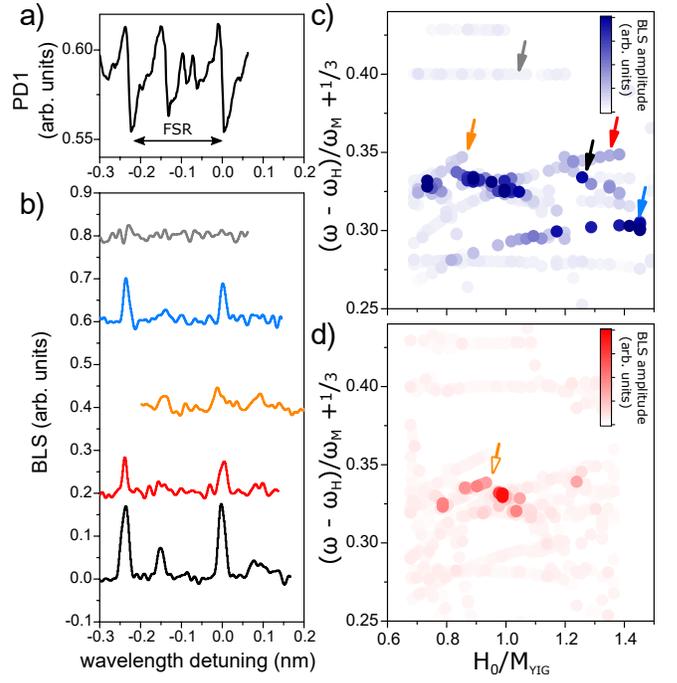}%
\caption{Optical measurements of BLS efficiency. (a) The reflected input optical power at photodiode I as a function of input laser wavelength detects the WGM resonances. The laser detuning ($x$-axis) is measured relative to one of the WGMs. Resonances corresponding to mode families $q=1,2$ are observed. The free spectral range for modes with $q=1$ is indicated. (b) BLS signal as a function of input laser wavelength for several magnons marked by arrows in (c) with matching color. (c) BLS intensity maxima of the $q=1$ WGM resonance (coded by color intensity) for each measured point of microwave frequency ($y$-axis, reduced frequency) and magnetic field ($x$-axis). (d) Same as (c), but with static magnetic field inverted.}%
\label{BLS}%
\end{figure}

The non-uniformity of the static magnetic field leads to microwave absorption of additional modes and normal mode splitting at degeneracies, in particular between $\{110\}$ and $\{200\}$. Nevertheless, the modes identified in Fig.\thinspace\ref{msm}(b) are easily recognized and labeled by the colored arrows. At higher magnetic fields, corresponding to the YIG sphere being closer to the permanent magnet, the increasing non-uniformity of the magnetic field further distorts the spectra.

Finally, we put the rutile coupling prism in place next to the YIG sphere. The  prism is a good microwave dielectric resonator, so that the spectra in Fig.\thinspace \ref{msm}(c) are affected by a large number of additional spurious resonances. These do not depend on the magnetic field and have a negative slope since the Larmor frequency has been subtracted. Despite this, the magnetostatic modes can still be clearly identified.

\begin{figure}%
\includegraphics[width=\columnwidth]{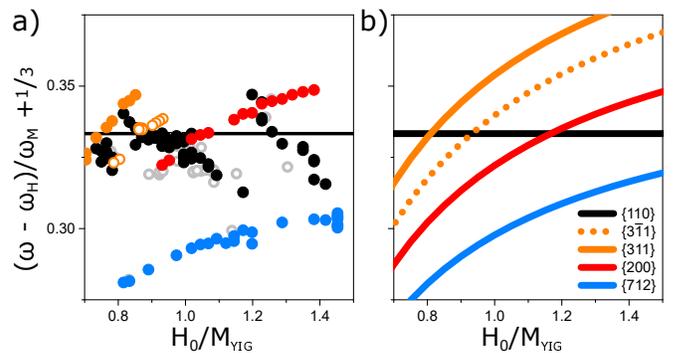}%
\caption{Indentification of BLS active magnetostatic modes. (a) Data points corresponding to those in Fig.\,\ref{BLS}(c) which are above a threshold set by the noise floor. The dot colors correspond to the assigned magnetostatic mode. Positive and negative magnetic fields are indicated by filled and open points, respectively. Points where mode identification is unclear are marked in gray. (b) Theoretical dispersion of the modes observed in (a).}%
\label{active}%
\end{figure}

\section{Brillouin light scattering}

Having identified the magnons that couple to the microwave antenna, we now discuss the optical measurements. We identify WGM resonances by varying the wavelength of the input light and measuring the reflected output in photodiode I, as shown in Fig.\,\ref{BLS}(a). Dips are seen for the $q=1$  mode family as well as a smaller peak for the $q=2$ mode family \cite{haigh_triple-resonant_2016}, where $q$ is the radial index of WGM defined above. 

We apply a microwave drive at several of the magnetostatic mode frequencies identified in Fig.\,\ref{msm}(c) and look for BLS. The polarization scattered light is spectrally resolved using a Fabry-P\'erot etalon to identify the anti-Stokes BLS and measured on photodiode II. For each microwave resonance we sweep the input laser wavelength, some of the spectra are plotted in Fig.\,\ref{BLS}(b). The peaks indicate enhanced BLS when the drive laser is resonant with a WGM.

A fit to the BLS peaks corresponding to the $q=1$ mode is used to extract the maximum BLS for each FMR frequency. The results are plotted as a function of reduced frequency and positive and negative magnetic field in Fig.\,\ref{BLS}(c) and Fig.\,\ref{BLS}(d),  respectively. In addition to the Kittel mode, several magnetostatic modes also generate BLS above the noise level.

We take the data from Fig.\,\ref{BLS}(c,d) and set a suitable noise-level threshold determined from a histogram of the measurement points. The measurement points with BLS above this threshold are plotted in Fig.\,\ref{active}(a). Closed (open) circles indicate measurements at positive (negative) magnetic field. Using their dispersion from Fig.\,\ref{msm}, we identify the magnetostatic mode associated with each of the points. The relevant calculated magnetostatic mode frequency dispersion are plotted for comparison in Fig.\,\ref{active}(b). For some points with negative field, mode identification has not been possible (gray points) due to the proximity of the overwhelming signal of the Kittel mode.

\section{Comparison with theory}

We are now in a position to compare the observed mode frequencies with the model calculations. In Sec.\,\ref{Sec:Theory}, we concluded that BLS should be observed for magnetostatic modes with odd $l_m$ and $m_{m}={+1}$(${-1}$) for positive (negative) magnetic field, respectively. In addition to the Kittel mode $\{110\}$, we observe the $\{311\}$ mode and, in the opposite field direction, the $\{3\bar{1}1\}$ mode, as expected. Additionally, we observe a signal for the $\{712\}$ mode. We do not observe BLS for the $\{220\}$, $\{330\}$, $\{320\}$ and several other magnetostatic modes identified in the microwave measurements, all conforming to the selection rules derived above.

On the other hand, the BLS by the $\{200\}$ mode contradicts the model predictions. This is likely caused by the non-uniformity of the applied magnetic field discussed in Sec.\thinspace \ref{Sec:MW_Exp}. The $m=0$ modes are particularly sensitive to inhomogeneities that break axial symmetry, as they are identical with spin-waves in the bulk material \cite{roschmann_properties_1977}. In theory, all that is required to allow BLS would be a small lateral shift in the $\{200\}$ mode function with respect to the center of the sphere. This is plausible given the magnetic field inhomogeneity. The axial symmetry breaking also allows resonant coupling to the ${110}$ mode \cite{roschmann_properties_1977}, which is evidenced in our microwave experiments as a normal mode splitting between the $\{200\}$ and $\{110\}$ modes (cf. Fig.\thinspace \ref{msm}(c) and Fig.\thinspace \ref{msm}(a)). Note that the BLS scattering from the {200} mode is still observed far from the anticrossing, indicating that this effect is not simply due to resonant admixing of the two mode functions (see Fig.\,4(c)). While the non-uniformity of the magnetic field complicates the interpretation, it does indicate that BLS by magnetostatic modes can be tailored by the application of controlled non-uniform magnetic fields.

For the $\{712\}$ mode, we measure similar BLS strength to the $\{110\}$ (see Fig.\,\ref{BLS}(b)). However, the microwave coupling to the $\{712\}$ is much weaker than that to the $\{110\}$. This can be seen from the fact that the ratio of the observed depth of the microwave resonances $\approx1/8$ (see Fig.\,\ref{msm}(c)), while the internal $Q$-factors are approximately equal \footnote{Note that the loaded $Q$-factor of the $\{110\}$ mode is significantly lower due to strong radiative damping via the antenna.}. Thus, the optomagnonic coupling must be stronger, in order that the BLS is comparable. This is consistent with calculations that show that the optomagnonic coupling for the $\{712\}$ mode is 3--4 times larger than for the $\{110\}$ mode.

\section{Conclusions}

In conclusion, we have measured cavity enhanced BLS from magnetostatic modes other than the uniform Kittel mode. We find reasonable agreement with the recently determined selection rules based on the axial rotational symmetry of the system \cite{Sharma_light_2017}. If microwave coupling to higher order modes can be optimized, the stronger optomagnonic coupling strength could be exploited. This offers a possible route to achieving larger microwave-to-optical conversion efficiency.

Our experimental results are in broad agreement with a recent paper \cite{osada_brillouin_2018} covering related experiments.

\section*{Acknowledgments}

We are grateful to Andreas Nunnenkamp and Mattias Weiler for useful discussions. This work was supported by the European Union’s Horizon 2020 research and innovation programme under grant agreement No 732894 (FET Proactive HOT), the Netherlands Organization for Scientific Research (NWO), and Grant-in-Aid for Scientific Research (Grant Nos. 25220910, 26103006) of the Japan Society for the Promotion of Science (JSPS).

The data plotted in the figures can be accessed at the Zenodo repository \cite{j_a_haigh_2018}.

\bibliography{bibliography}

\end{document}